\documentstyle[amssymb,secnumtab]{fbssuppl}

\def\be{\begin{eqnarray}}
\def\ee{\end{eqnarray}}
\def\bq{\begin{equation}}
\def\eq{\end{equation}}
\def\pr{Phys. Rev. }
\def\np{Nucl. Phys. }\def\pl{Phys. Lett. }

\title{Nucleon structure functions in a constituent quark scenario
\thanks{Supported in part by DGICYT-PB94-0080 
and TMR programme of the European Commission ERB FMRX-CT96-008;
based on two talks given at: 1) Workshop ``N$^*$ Physics and non perturbative
QCD'', Trento, Italy, May 18-29 1998, to be published in Few Body Systems, 
Suppl.;
2) Conference ``Fev Body 16$^{th}$'', Autrans, France,
June 1-6, 1998,  to be published in Few Body Systems, Suppl.}}
\author{{Sergio Scopetta}$^a$, 
Vicente Vento $^{a,b}$,
and Marco Traini $^c$  }
\institute{
$^a$ Departament de Fisica Te\`orica, Universitat de Val\`encia \\
46100 Burjassot (Val\`encia), Spain; 
\\ 
$^b$ Institut de F\'{\i}sica Corpuscular, Consejo Superior de 
Investigaciones Cient\'{\i}ficas; 
\\
$^c$ Dipartimento di Fisica, 
Universit\`a di Trento, I-38050 Povo (Trento), Italy, and 
INFN, Gruppo Collegato di Trento.}

\begin{document}

\maketitle
\begin{abstract}
Using a simple picture of the constituent quark as a
composite system of point-like partons, we construct the 
polarized parton distributions by
a convolution between constituent quark momentum distributions
and constituent quark structure functions.
We achieve good 
agreement with experiments in the unpolarized as well as in the
polarized case, though a good description of the recent polarized
neutron data requires the introduction of one more parameter.
When our results are compared with similar calculations using non-composite 
constituent quarks, the accord with the experiments of the present scheme 
is impressive. We conclude that DIS data are consistent with a low energy 
scenario dominated by composite constituents of the nucleon. 

\end{abstract}


At low energies, 
the so called naive quark model accounts for a large number of experimental
observations. 
At large energies,
$QCD$ sets the framework for an
understanding of the Deep Inelastic Scattering (DIS)
phenomena beyond the Parton
Model. However, the perturbative approach to $QCD$
does not provide absolute values for the observables. The description based
on the Operator Product Expansion ($OPE$) and the $QCD$ evolution requires the
input of non-perturbative matrix elements. We have developed an approach which 
uses model calculations for the latter ingredients \cite{Traini}. 
Moreover, in order to relate the constituent quark with the current partons of
the theory, a procedure, hereafter called ACMP, has been applied
\cite{Altarelli1,Scopetta}.
Within this approach, constituent quarks are effective particles 
made up of point-like partons (current quarks (antiquarks) and gluons), 
interacting by a residual interaction described as in a quark model. 
The hadron structure
functions are obtained by a convolution of the constituent quark model
wave function with the constituent quark structure function. 
This idea has been recently used to estimate the pion structure function
\cite{Altarelli2}.
We summarize here our application to the unpolarized
\cite{Scopetta} and polarized \cite{pol} DIS off the nucleon. 
It will be found that $DIS$ data are consistent with a 
low energy scenario dominated by composite
constituents. 


In our picture the constituent quarks are
themselves complex objects whose structure functions are described by a set of
functions $\Phi_{ab}$ that specify the number of point-like partons 
of type $b$, which are present in the constituents of type $a$ with fraction 
$x$ of its total momentum \cite{Altarelli1,Scopetta}. In general $a$ and $b$ 
specify all the relevant quantum numbers of the partons, i.e., flavor and 
spin. Let us discuss first the unpolarized case for the proton 
\cite{Scopetta}.

The functions describing the nucleon parton distributions omitting spin degrees
of freedom are expressed in terms of the independent $\Phi_{ab}(x)$ and of the
constituent probability distributions $u_0$ and $d_0$, at the hadronic scale
$\mu_0^2$ \cite{Traini}, as
\bq
f(x,\mu_0^2) = \int_x^1 \frac{dz}{z}[u_0(z,\mu_0^2)\Phi_{uf}
(\frac{x}{z},\mu_0^2) +
d_0(z,\mu_0^2)\Phi_{df}(\frac{x}{z},\mu_0^2)]
\eq
where $f$ labels the various partons, i.e., valence quarks ($u_v,d_v$), sea
quarks ($u_s,d_s,s$), sea antiquarks ($\bar{u},\bar{d},\bar{s}$) 
and gluons $g$.
The different types and functional forms of the structure functions for the
constituent quarks are derived from three very natural assumptions
\cite{Altarelli1}:
{\it i)} 
The point-like partons are the quarks, 
antiquarks and gluons described by $QCD$;
{\it ii)} Regge behavior for $x\rightarrow 0$ and duality ideas;
{\it iii)} invariance under charge conjugation and isospin.

These considerations define the following
structure functions \cite{Altarelli1}

\bq
\Phi_{qf}({x}, \mu_0^2)
= C_f
 x^{a_f} (1-x)^{A-1}~,
\label{csf1}\eq
where $f=q_v,\,q_s,\,g$ for the valence quarks, the sea and the gluons,
respectively.
Regge phenomenology suggests:  $ a_{q_v} = -0.5 $ ($\rho$ meson
exchange) and  $a_{q_s} = a_g = -1$ ($pomeron$ exchange).
The other ingredients of the formalism, i.e., 
the probability distributions for 
each constituent quark, are defined according to the procedure of ref.
\cite{Traini} and shown in \cite{Scopetta}.
Our last assumption relates to the hadronic scale $\mu_0^2$, i.e., 
that  at which the constituent quark structure is defined. We choose 
$\mu_0^2=0.34$ GeV$^2$, as defined in Ref. \cite{Traini}, 
namely by fixing the momentum carried by the various partons. 
This choice  of the hadronic scale determines all the parameters 
except one, which is fixed through 
the data \cite{Scopetta}.
To complete the process, the above input 
distributions are NLO-evolved in the DIS scheme to the experimental
scale, where they are compared with the data.

We next generalize our previous discussion to the polarized parton
distributions. 
As it is explained in ref. \cite{pol},
using $SU(6)$ (spin-isospin) symmetry and other
reasonable simplifying assumptions,
it can be shown 
that 

\bq
\Delta f(x,\mu_0^2) = \int_x^1 \frac{dz}{z}[u_0(z,\mu_0^2)
\Delta \Phi_{uf} (\frac{x}{z},\mu_0^2) +
d_0(z,\mu_0^2)\Delta \Phi_{df}(\frac{x}{z},\mu_0^2)]~,
\eq

\begin{figure}[h]
\vspace{4.cm}
\includegraphics{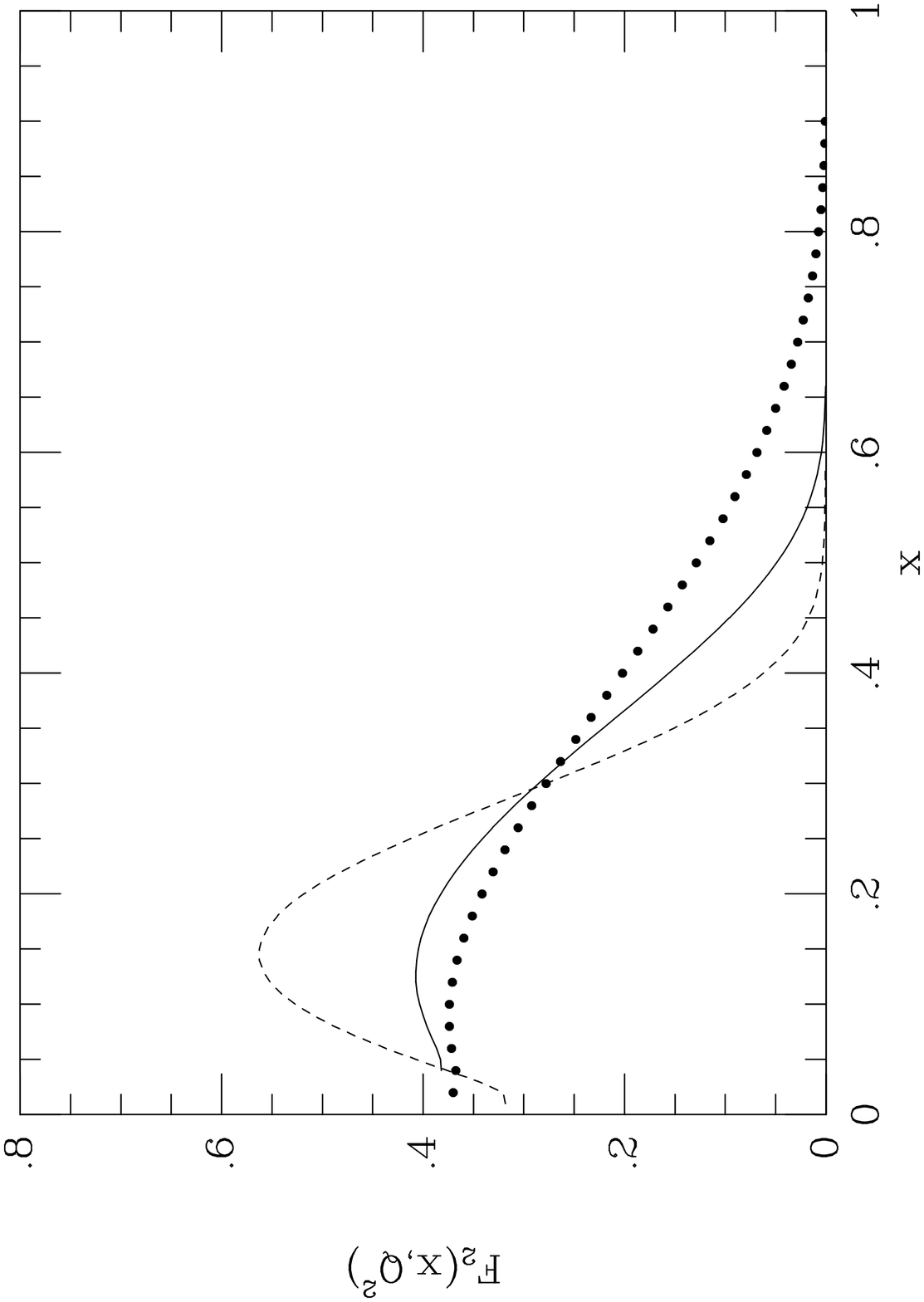}
\includegraphics{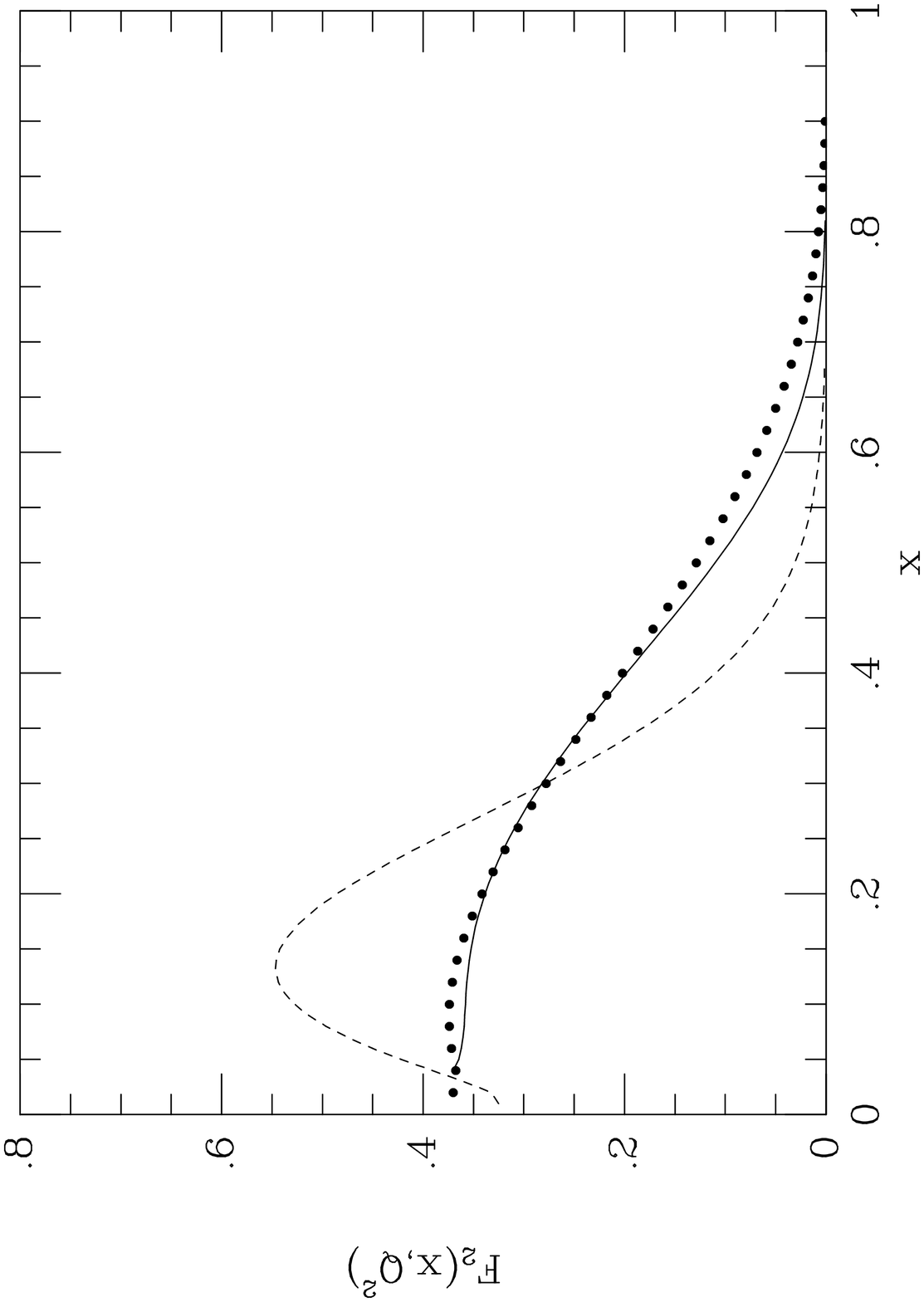}
\caption{The proton $F_2(x,Q^2)$, obtained
by NLO-evolution to $Q^2=10$ GeV$^2$ (full), compared to the data
(dots) \cite{emc}. 
The result which would be obtained disregarding the constituent
structure is also shown (dashed). Left (right) panel: 
constituent wave functions form ref. \cite{ik} (ref. \cite{iac}).}
\end{figure}
\noindent where $f$ labels the various partons;
it means that the $ACMP$ procedure can
be extended to the polarized case just by introducing three additional
structure functions for the constituent quarks: $\Delta \Phi_{q q_v}$, $\Delta
\Phi_{q q_s}$ and $\Delta \Phi_{q g}$.
In order to determine them we add two
minimal assumptions:
{\it iv)} factorization: $\Delta \Phi$ cannot depend upon the quark 
model used;
{\it v)} positivity: the constraint $\Delta \Phi \leq
\Phi $ is saturated for $x = 1$.
In such a way we determine completely 
the $\Delta \Phi$'s.
In fact, the $QCD$ partonic picture, Regge behavior and duality imply that
\bq
\Delta\Phi_{q f} = {\Delta C_f  {x^{- \Delta a_f}} (1-x)^{\Delta A_f - 1} }
\label{dcsf}
\eq
and $ -\frac{1}{2}< \Delta a_f < 0$, for all 
$f= q_v, q_s, g$, as allowed by dominant 
exchange of the $A_1$ meson trajectory \cite{abfr}. 
Moreover, the assumption that the positivity restriction 
is saturated for $x=1$, in the spirit of ref. \cite{kaur}, 
implies that the $\Phi's$ and the $\Delta \Phi's$
have the same large $x$ behavior, and that $\Delta C_f = C_f$,
(the latter being introduced in (\ref{csf1})); 
it means that the
partons which carry all of the momentum also carry all of the polarization.
Let us stress that the change between the polarized
functions and the unpolarized ones comes only from Regge behavior;
as a matter of fact, it turns out that, {\it except for the exponent
$\Delta a_f$} shown above, the $\Delta \Phi$'s, Eq. (\ref{dcsf}),
are given by the unpolarized functions, Eq. (\ref{csf1}). 
The other ingredients, i.e., 
the polarized distributions for 
each constituent quark, are defined according to the procedure of
ref. \cite{Traini} and they are shown in ref. \cite{pol}.
Finally, the parton distributions at the hadronic scale 
are evolved to the experimental
scale by performing a NLO evolution in the AB scheme \cite{abfr}.
Results are shown in Figs. 1 and 2.  Fig. 1 refers to the unpolarized case.
The structure function $F_2(x,Q^2)$, obtained 
evolving the parton distributions 
Eq. (1), calculated using Eq. (\ref{csf1})
for the $\Phi_{qf}$'s
and two different models for $u_o$ and $d_o$,
describes successfully the data. The agreement becomes impressive
if compared with a similar calculation with non-composite constituents. 
\begin{figure}[h]
\vspace{4.0cm}
\includegraphics{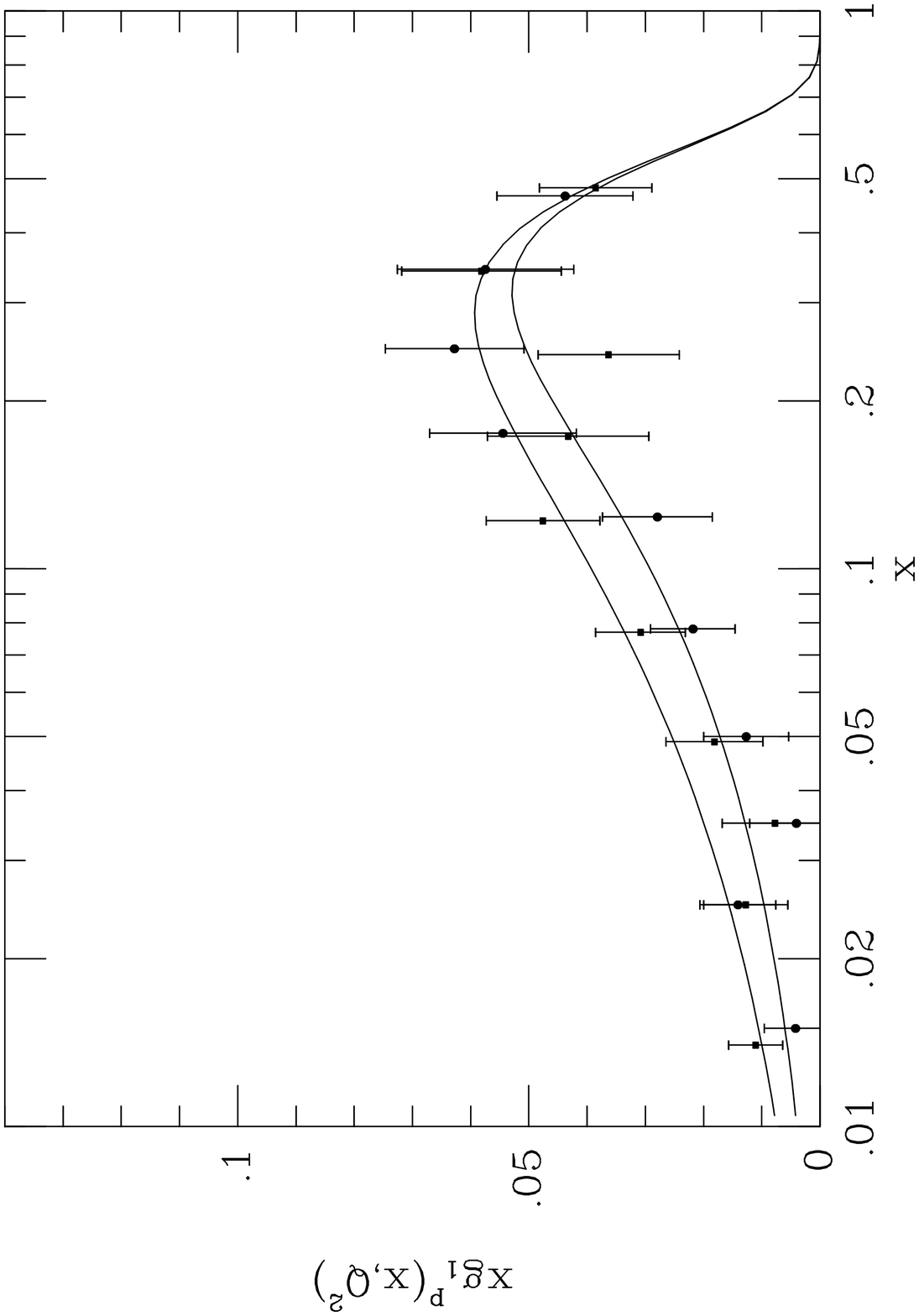}
\includegraphics{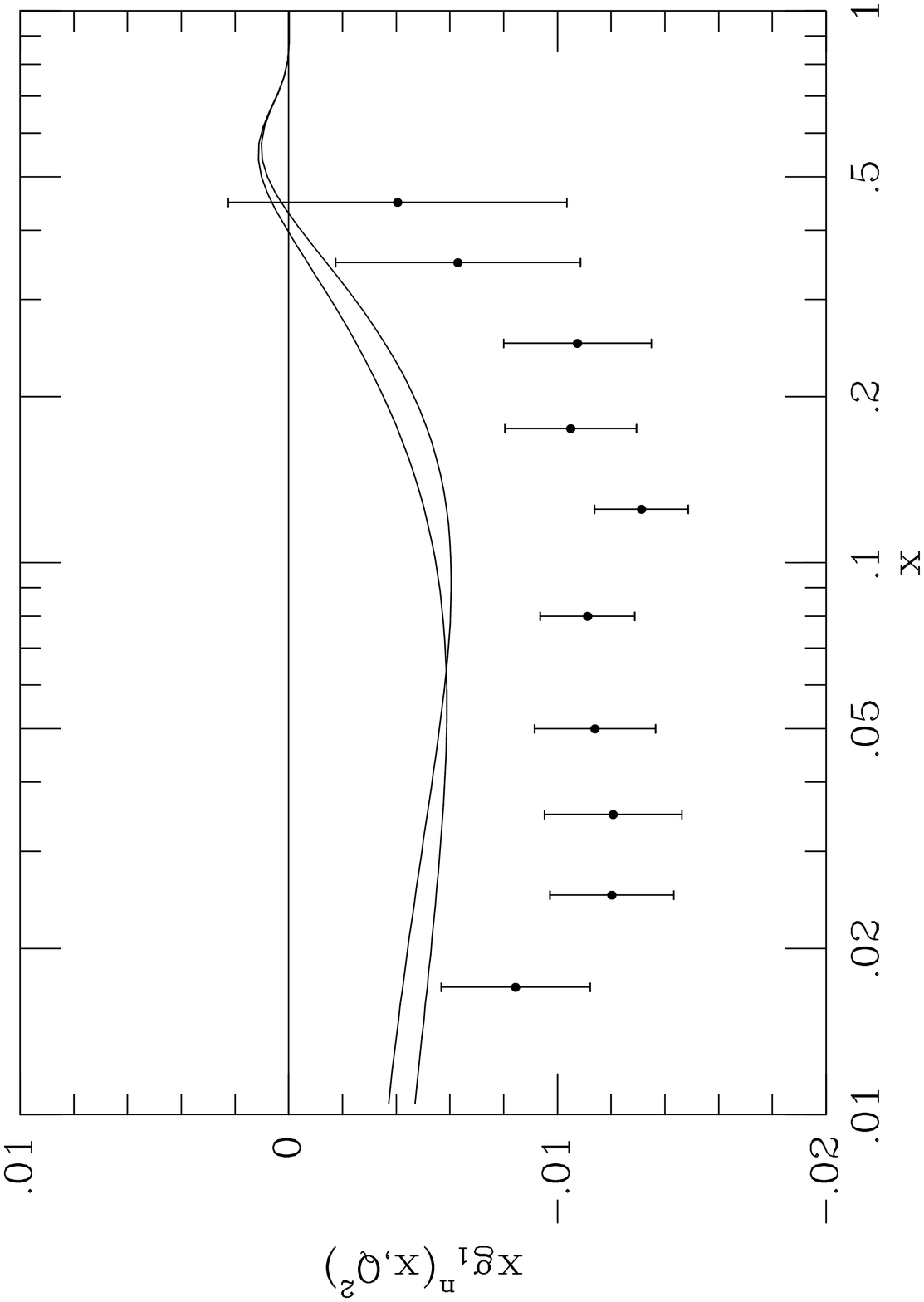}
\caption{
Left (Right): $xg_1(x,Q^2)$ for the proton
(neutron) evolved at NLO to
$Q^2= 10\,(5)$ GeV$^2$, for the two extreme Regge behaviors
mentioned in the text (full curves). 
The wave functions used are from ref. \cite{iac}.
The data \cite{emc} are shown for comparison.}
\end{figure}

In the polarized case, it is found \cite{pol} that the constituent
structure functions Eq. (\ref{dcsf}) give a good result for
the proton, but they fail in reproducing the recent
precise neutron data. 
This is to be ascribed 
to our naive input for the 
sea and to the 
symmetry for the $u$ and $d$ quarks \cite{pol}. In particular,
it has been shown that, by redefining the sea $\Delta \Phi$, changing
{\it only one} parameter so that the experimental 
sea polarization is recovered, also the neutron is
rather well described. Fig. 2 refers to this last scenario.
The procedure is also able
to predict successfully several observables, such as the nucleon axial
charges \cite{pol}. It should be noticed that in this framework
the {\it spin crisis}, as initially presented, does not arise.  

Summarizing, low energy models seem to be consistent
with DIS data when a structure for the constituent is introduced.
The crucial role played by the sea in the polarized case,
as well as the implementation of Chiral Symmetry Breaking 
in our procedure, have to be more deeply investigated. 
It will be the subject of future work.

\end{document}